# Kaluza-Klein FRW type dark energy model with a massive scalar field

D.R.K.Reddy[1], K. Deniel Raju[1,2]

[1]*Department of Applied Mathematics, Andhra University, Visakhapatnam-530003, India.*
[2]*Department of Mathematics, ANITS (A), Visakhapatnam-531162, India*
Email : reddy_einstein@yahoo.com

**Abstract:** In this investigation we discuss the dynamical aspects of Kaluza-Klein FRW type cosmological model in the presence of dark energy fluid and an attractive massive scalar field. The field equations are solved using a power law between the average scale factor and the scalar field to reduce the mathematical complexity. We have presented the corresponding dark energy model. Important cosmological parameters like equation of state (EoS) parameter, the deceleration parameter, the density parameter and the jerk parameter are evaluated. The physical behavior of the parameters is also discussed.

## 1. Introduction

Type 1a Supernova [1-2] and Cosmic Microwave Background Radiation (CMBR)[3-4] confirmed that the observable universe is spatially flat and accelerating. The reason for this accelerating expansion of the universe, it is believed, is an exotic fluid known as dark energy (DE) with a large negative pressure. However, DE is still a big mystery. Cosmological constant was considered to be the simplest candidate for DE. But this faces the cosmological constant problem. Subsequently, to explain DE, in a more satisfactory way, several DE models have been proposed from time to time. It is said that another significant way to account for the cosmic acceleration is to modify general relativity and study scalar field or quintessence models in modified theories of gravitation because of the fact that they represent DE models. By modifying the Einstein-Hilbert action several modified theories have been proposed and DE models have been studied. Significant modified theories are scalar-tensor theories of gravitation proposed by Brans and Dicke [5] and Saez Ballester [6] and modified theories like f(R) and f(R,T) theories gravitation[7-8]. A comprehensive review of DE models and modified theories of gravity has been presented by Padmanabham [9], Copeland et al. [10] and Bamba et al. [11].

With a view to unify gravity with other gauge interactions, there has been a lot of interest in the study of higher dimensional space-time. It is believed that at the early stage of evaluation of the universe, it might have been preceded by a multidimensional stage. However, as time evolves standard four dimensions expand while the extra dimensions shrink due to compactification. This study has attracted many researchers to investigate cosmological models in higher dimensions [12-14]. Kaluza [15] and Klein [16] developed five dimensional general relativity in an attempt to unify gravity and electromagnetism. Many authors have studied several cosmological models using five or more dimensions. [17-22];

It is well known that scalar field models play a vital role in explaining DE models. The most significant scalar fields are Brans-Dicke [5], Saez-Ballester [6] and Barber's [23] scalar fields. There have been several investigations, in literature, on Brans-Dicke and Saez-Ballester DE models. Note worthy among them are DE models obtained by Rao et al. [24], Reddy [25], Reddy et al. [26], Kiran et al. [27], Reddy et al. [28], Reddy et al. [29], Rao et al. [30]. DE models in the presence of scalar meson fields are also important in explaining late time acceleration of the universe. Scalar meson fields are of two types - zero mass (mass less) scalar fields and attractive massive scalar fields. In recent years, there has been lot of interest in constructing DE models in the presence of zero mass and massive scalar fields. Reddy et al. [31], have obtained anisotropic dark energy cosmological model coupled with mass less scalar meson fields in general relativity while Reddy and Ramesh [32] discussed five dimensional DE model in the presence of zero mass scalar meson fields. Bianchi type DE model in the presence of massive scalar fields have been discussed by Reddy [33], Naidu [34], Aditya and Reddy [35], Naidu et al. [36] and Singh and Rani [37].

In this paper, we discuss the dynamical aspects of Kaluza- Klein FRW type DE model in the presence of massive scalar field. The motivation for this work comes from the above discussion. Sec. 2 is devoted to the derivation of gravitational field equations with the aid of FRW type Kaluza-Klein metric with DE fluid and massive scalar field as source. In Sec. 3, we solve the field equations and present the corresponding cosmological model. Significant cosmological parameters of the model are computed and their physical discussion is given in Sec. 4. In the last section we give the summary and conclusions.

**2. Field Equations**

The space-time given by the metric

$$ds^2 = dt^2 - a^2(t)\left[\frac{dr^2}{1-kr^2} + r^2(d\vartheta^2 + \sin^2\vartheta d\phi^2) + (1-kr^2)d\psi^2\right] \quad (1)$$

defines the geometry of FRW type Kaluza-Klein cosmology. Here $a(t)$ is the scale factor, $k = -1, 0, 1$ is the curvature parameter for spatially closed, flat and open universe respectively.

Einstein's field equations with anisotropic DE fluid and an attractive massive scalar field $\varphi$ as source of gravity are given by

$$R_{ij} - \frac{1}{2}g_{ij}R = -(T_{ij}^{(de)} + T_{ij}^{(\phi)}) \quad (2)$$

Here the energy momentum tensor for anisotropic DE fluid is given by

$$T_{ij}^{(de)} = (\rho_{de} + p_{de})u^i u_j - p_{de}g_{ij} \quad (3)$$

and $T_{de}^{(\phi)}$ corresponding to scalar field is given by

$$T_{ij}^{(\phi)} = \varphi_{,i}\varphi_{,j} - \frac{1}{2}(g_{ij}\varphi_{,k}\varphi^{,k} - M^2\varphi^2) \quad (4)$$

where M is the mass of the scalar field.

Also the velocity $u^i$ satisfies

$$\begin{aligned} u^i u_i &= 1 \\ u^i u_j &= 0 \\ \varphi &= \varphi(t) \end{aligned} \quad (5)$$

The other symbols have their usual meaning and the scalar field $\varphi$ satisfies the Klein-Gordan equation.

$$g^{ij}\varphi_{;ij} + M^2\varphi = 0 \quad (6)$$

Here a coma and a semicolon denote ordinary and covariant differentiation.

The energy momentum tensor of DE fluid can be parameterized as

$$T_{ij}^{(de)} = \text{diag}[1, -\omega_{de}, -(\omega_{de} + \gamma), -(\omega_{de} + \delta) - (\omega_{de} + \beta)]\rho_{de} \quad (7)$$

Where $\omega_{de} = \frac{p_{de}}{\rho_{de}}$ is the equation of state (EoS) parameter of DE and $\gamma, \delta, \beta$ are the skewness parameters along $\vartheta, \phi, \psi$ directions respectively.

Now using co moving coordinates and equations (3) - (5) and (7) the field equations (2) for the metric (1) can be written, explicitly, as

$$6\frac{\dot{a}^2}{a^2} + \frac{6k}{a^2} = \rho_{de} + \frac{\dot{\varphi}^2}{2} + \frac{M^2\varphi^2}{2} \tag{8}$$

$$3\left(\frac{\ddot{a}}{a} + \frac{\dot{a}^2}{a^2} + \frac{k}{a^2}\right) = -\rho_{de}\omega_{de} - \frac{\dot{\varphi}^2}{2} + \frac{M^2\varphi^2}{2} \tag{9}$$

$$3\left(\frac{\ddot{a}}{a} + \frac{\dot{a}^2}{a^2} + \frac{k}{a^2}\right) = -\rho_{de}(\omega_{de} + \gamma) - \frac{\dot{\varphi}^2}{2} + \frac{M^2\varphi^2}{2} \tag{10}$$

$$3\left(\frac{\ddot{a}}{a} + \frac{\dot{a}^2}{a^2} + \frac{k}{a^2}\right) = -\rho_{de}(\omega_{de} + \delta) - \frac{\dot{\varphi}^2}{2} + \frac{M^2\varphi^2}{2} \tag{11}$$

$$3\left(\frac{\ddot{a}}{a} + \frac{\dot{a}^2}{a^2} + \frac{k}{a^2}\right) = -\rho_{de}(\omega_{de} + \beta) - \frac{\dot{\varphi}^2}{2} + \frac{M^2\varphi^2}{2} \tag{12}$$

The conservation equation for matter energy tensor takes the form

$$\dot{\rho}_{de} + 4\frac{\dot{a}}{a}(1 + \omega_{de})\rho_{de} = 0 \tag{13}$$

and the Klein-Gordon equation becomes

$$\ddot{\varphi} + 4\frac{\dot{a}}{a}\dot{\varphi} + M^2\varphi = 0 \tag{14}$$

where an overhead dot denotes differentiation with respect to time t

From Equations (9) - (12) we have

$$\gamma = \delta = \beta = 0 \tag{15}$$

so that the following field equations to solve [Eq. (13) being conservation equation]

$$6\frac{\dot{a}^2}{a^2} + \frac{6k}{a^2} = \rho_{de} + \frac{\dot{\varphi}^2}{2} + \frac{M^2\varphi^2}{2} \tag{16}$$

$$3\left(\frac{\ddot{a}}{a} + \frac{\dot{a}^2}{a^2} + \frac{k}{a^2}\right) = -\rho_{de}\omega_{de} - \frac{\dot{\varphi}^2}{2} + \frac{M^2\varphi^2}{2} \tag{17}$$

$$\ddot{\varphi} + 4\frac{\dot{a}}{a}\dot{\varphi} + M^2\varphi = 0 \tag{18}$$

## 3. Solution and the model

Now the field equations (16) - (18) are a system of three independent equations in four unknowns $a, \rho_{de}, \omega_{de}$ and $\varphi$. Hence to obtain a deterministic model and to avoid mathematical complexity we use

$$4\frac{\dot{a}}{a} = -\frac{\dot{\varphi}}{\varphi} \tag{19}$$

which amounts to a power law between the average scale factor $a(t)$ and the scalar field $\varphi$ (Singh and Rani [30]; Naidu et al. [28]).

Now from Eqs. (18) and (19) we have on integration

$$\varphi = \exp\left(\varphi_0 t - \frac{M^2 t^2}{2} + \varphi_1\right) \tag{20}$$

Now from Eqs. (20) and (19) we obtain

$$a(t) = \exp\left(\frac{\frac{M^2 t^2}{2} - \varphi_0 t - \varphi_1}{4}\right) \tag{21}$$

Using Eq. (21) in Eq. (1) we obtain the FRW type Kaluza-Klein DE model

$$ds^2 = dt^2 - \exp\left(\frac{\frac{M^2 t^2}{2} - \varphi_0 t - \varphi_1}{2}\right)\left[\frac{dr^2}{1-kr^2} + r^2(d\theta^2 + \sin^2\theta d\varphi^2) + (1-kr^2)d\psi^2\right] \tag{22}$$

with the massive scalar field in the model given by Eq. (20).

## 4. Cosmological parameters in the model

The DE model given by Eq. (22) has the following cosmological parameters which are essential for the discussion of dynamical aspects of the model.

The spatial volume and the average scale factor are given by

$$V = a^3(t) = \left[\exp\left(\frac{\frac{M^2 t^2}{2} - \varphi_0 t - \varphi_1}{4}\right)\right]^3 \tag{23}$$

Average Hubble parameter is

$$H = \frac{\dot{a}}{a} = \frac{1}{4}\left(\frac{M^2 t - \varphi_0}{4}\right) \tag{24}$$

The scalar expansion $\theta$ is

$$\theta = 4H = \left(\frac{M^2 t - \varphi_0}{4}\right) \tag{25}$$

The deceleration parameter is

$$q = -\left(1 + \frac{16 M^2}{(M^2 t - \varphi_0)^2}\right) \tag{26}$$

From Eqs.(16),(20) and (21) we obtain DE density as

$$\rho_{de} = \frac{3}{8}(M^2 t - \varphi_0)^2 - \frac{1}{2}\left[\exp(2\varphi_0 t - M^2 t + 2\varphi_1)\right]\left[(M^2 t - \varphi_0)^2 - M^2\right] + 6k \exp\left(\frac{2\varphi_0 t - M^2 t - 2\varphi_1}{4}\right) \tag{27}$$

From (17),(20) and (21) we get the EoS parameter as

$$\omega_{de} = -\frac{1}{\rho_{de}}\left[\frac{3}{16} + \frac{1}{2}\exp(2\varphi_0 t - M^2 t + 2\varphi_1) + M^2\left(\frac{3}{4} - \frac{1}{2}\exp(2\varphi_0 t - M^2 t + 2\varphi_1)\right) + 3k\left(\frac{\exp(2\varphi_0 t - M^2 t + 2\varphi_1)}{4}\right)\right] \tag{28}$$

where $\rho_{de}$ is given by Eq.(28).

The jerk parameter in the model is found to be

$$j(t) = 1 + \frac{80 M^2}{(M^2 t - \varphi_0)^2} + \frac{528 M^2}{(M^2 t - \varphi_0)^4} \tag{29}$$

**Physical discussion:** Eq.(22) describes the FRW type Kaluza-Klein DE model in the presence of massive scalar field in Einstein's theory of gravitation. It can be seen that the model has no initial singularity. The spatial volume increases with time from a finite volume at t=0. This shows that the universe represented by our model is expanding with time. It may be observed that $H, \theta, \rho_{de}, \omega_{de}$ diverse as t approaches infinity and w e get a finite value at t=0. It may also be observed that DE density is always positive through out the evolution of the universe. Also we have EoS parameter as function of cosmic time and $\omega_{de} > -1$ which implies that our model is quintessence model. It is seen that q=-1 at late times showing that our model is accelerating. Since q=- 1 and j=1 at late times we can conclude that there is a smooth transition from deceleration to accelerating phase of the universe. Also as t tends to infinity we have $\Lambda CDM$ model of the universe. Thus our model is in good agreement with the recent scenario of accelerating universe

## 5. Conclusions

This paper is to devoted to the investigation of Kaluza-Klein FRW type cosmological model in the presence of DE fluid it is greater with an attractive massive scalar field in general relativity. We have found a deterministic model of the universe using a power law between the average scale factor of the universe and the scalar field. we have als determined all the significant cosmological parameters of our model and a physical discussion relevant to the scenario of modern cosmology is presented. We have the following results:

- Our model is non singular and expanding from a finite volume.
- The Hubble parameter, the scalar expansion diverge as t tends to infinity
- The DE density is alwas positive and tends to infinity from a finite value
- The EoS parameter is a function of cosmic time and at late times $\omega_{de} > -1$ which implies that our model is a quintessence model which should infact be the case as our model is a scalar model.
- Thus or model agrees with recent cosmological data[1-2].
- Since at late times $q = -1$ our model is accelerating and since j=1 at late times our model exhibits a transition from decelerated phase to an accelerating phase.